\documentclass{mem}
\usepackage{natbib}
\usepackage{txfonts}
\usepackage{balance}
\usepackage{graphicx}
\usepackage[a4paper,breaklinks]{hyperref}

\idline{75}{282}
\begin{document}

\title{X-ray pulsars: a review}
   \subtitle{}
\author{I. \, Caballero\inst{1} \and J.\, Wilms\inst{2}}

\offprints{I. Caballero, \email{isabel.caballero@cea.fr}}
 
\institute{
DSM/IRFU/SAp- UMR AIM (7158), CNRS/CEA, University Paris Diderot, 91191 Gif-sur-Yvette, France
\and
Dr. Karl Remeis Sternwarte \& Erlangen Center for Astroparticle Physics, Sternwartstr. 7, 96049 Bamberg, Germany}

\authorrunning{Caballero,~I., Wilms,~J.}

\titlerunning{X-ray pulsars: a review}

\abstract{Accreting X-ray pulsars are among the most luminous objects in the X-ray sky. 
In highly magnetized neutron stars ($B\sim10^{12}$\,G), the flow of matter is dominated 
by the strong magnetic field. The general properties of accreting X-ray binaries are presented, 
focusing on the spectral characteristics of the systems. The use of cyclotron lines 
as a tool to directly measure a neutron star's magnetic field  
and to test the theory of accretion are discussed. We conclude with the current and 
future prospects for accreting X-ray binary studies. 
}
\maketitle{}

\section{Introduction}

Accreting X-ray binaries were discovered in the 70's (\citealt{giacconi71_1}, 
\citealt{tananbaum72}) and are among the most luminous objects in the X-ray sky. 
Forty years after their discovery, more than 400 X-ray binaries have been observed, and 
there is today a deeper knowledge of the physical processes involved in the accretion of 
matter onto the compact object. In X-ray binaries, 
matter is accreted from a donor star onto a compact object 
(white dwarf, neutron star or black hole).
X-ray emission originates as a result of the conversion of the gravitational energy of the accreted 
matter into kinetic energy. 
According to the nature of the donor star, X-ray binaries can be classified 
as High-Mass X-ray Binaries (hereafter HMXBs) or Low-Mass X-Ray Binaries 
(hereafter LMXBs). Typically, HMXBs have young optical companions of spectral type O or 
B and mass $M\gtrsim5M_{\odot}$, and high magnetic fields 
$B\sim10^{12}\,$$\mathrm{G}$. LMXB systems have older optical companions, with 
masses in general $M\leq 1\,M_{\odot}$, and lower magnetic fields 
$B\sim10^{9-10}\,$G. X-ray binaries are numerous objects in the Galaxy, with 
114 HMXBs in the Galaxy and 128 HMXBs in the Magellanic Clouds, 
and 187 LMXBs in the Galaxy and Magellanic Clouds \citep{liu05,liu06,liu07}. 
In accreting neutron stars, due to the inclination of the 
neutron star's rotational axis with respect to the magnetic axis, the X-ray emission appears 
pulsated to a distant observer. \\
\indent In this review, we concentrate on the properties of X-ray binaries with a highly 
magnetized neutron star ($B\sim10^{12}\,$$\mathrm{G}$). The basic aspects of accreting 
X-ray pulsars are described, and the current status of the observations and theory of these systems is presented.

\section{Accretion onto the neutron star}

The accretion of matter in X-ray binaries can take place via Roche lobe overflow or via direct 
wind accretion. Roche Lobe overflow takes place when the donor star evolves, expands and 
fills its Roche lobe. The matter exceeding the Roche lobe is no longer gravitationally bound 
to the star and can be accreted by the compact object through the inner Lagrange point. The 
accreted matter has a large amount of angular momentum, and therefore forms 
an accretion disk around the compact object instead of being accreted directly. This typically 
occurs in LMXBs. 

Wind accretion takes place in binary systems with a O or B donor star, namely HMXBs. These stars 
have very strong stellar winds, with a mass loss rate that can be as high as 
$\dot{M}\sim10^{-4}-10^{-6}\,M_{\odot}/\mathrm{yr}$. The compact object typically orbits the donor star  
at a distance of less than one stellar radius, and therefore the compact 
object is deeply embeded in the stellar wind (e.g., Vela X-1, \citealt{nagase86}). 

A large fraction of X-ray binaries (about 60\% of the HMXBs) falls in the category of Be/X-ray binary 
systems. These systems have quite eccentric orbits, and their X-ray activity is related to the 
presence of a disk of material that surrounds the equator of the Be star. They can have giant (type II) 
or normal (type I) outbursts, and long quiescence periods. See \cite{reig2011} for a recent 
review of Be/X-ray binaries. 

Independently of the way matter is transferred from the donor star to the compact object, at a  
distance close to the compact object the flow of matter is dominated by the strong magnetic field 
of $B\sim10^{12}\,$G. The infalling matter follows the magnetic field lines, forming accretion 
columns on the neutron star magnetic poles. The accreting plasma couples to the magnetic field lines 
at the Alfv\'{e}n radius 
$r_{\mathrm{mag}}=2.9\times10^{8}M_{1}^{1/7}R_{6}^{-2/7}L_{37}^{-2/7}\mu^{4/7}\mathrm{cm}$, 
where $M_{1}$ is given in units of $1\,M_{\odot}$, $R_{6}$ is given in units of 
$10^{6}\,\mathrm{cm}$ and $L_{37}$ is the luminosity in units of 
$10^{37}\,\mathrm{erg}\,\mathrm{s}^{-1}$. For typical neutron star parameters, 
$r_{\mathrm{mag}}\sim1800\,$km. Accretion can take place as long as the Alfv\'{e}n radius is 
smaller than the co-rotation radius $r_{\mathrm{co}}=\left(\frac{GM}{\omega^{2}}\right)^{1/3}$ 
(defined as the radius at which the angular velocity of the magnetosphere $\omega r$ and the 
Keplerian velocity of the disk  $\sqrt{GM/r}$ are equal). If this is not the case, the 
propeller effect can inhibit the accretion \citep{illarionov_sunyaev_1975}. Such an effect could 
for instance explain the off-states observed in Vela X-1 \citep{kreykenbohm08}. 

The luminosity produced by an accreting object is given by $L=GM\dot{m}/R$, where $M$, $R$  
are the mass and radius of the compact object, and $\dot{m}$ is the mass accretion rate. 
As will be further discussed in Sec.~\ref{sec:crsf}, there is a limit on the  luminosity 
that an accreting body can have, the Eddington luminosity, obtained by balancing gravitational 
force and radiation pressure. For a spherically symmetric accretion and a steady flow, the 
Eddington luminosity is 
$L_{\mathrm{Edd}}=\frac{4\pi GMm_{\mathrm{p}}c}{\sigma_{\mathrm{T}}}\approx 1.3\times10^{38}\frac{M}{M_{\odot}}\mathrm{erg}\,\mathrm{s}^{-1}$. 

\section{Formation of the spectral continuum} 

\begin{figure}
\resizebox{\hsize}{!}{\includegraphics[clip=true]{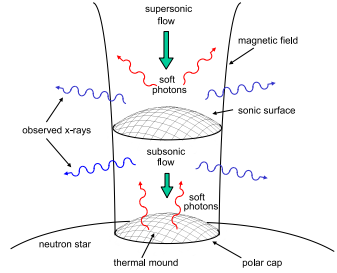}}
\caption{
\footnotesize
Schematic view of the accretion column. Seed photons are 
created along the column via bremstrahlung and cyclotron emission, together with black body 
seed photons emitted from the surface of the thermal mound. (Figure 1 from \cite{becker_wolff_07}, 
reproduced by permission of the AAS).
}
\label{fig:bw_1}
\end{figure}

The X-ray spectrum is typically well described by a power law with an exponential cutoff. 
Numerous authors have made attempts to derive the shape of X-ray pulsar
spectra analytically or numerically (e.g., \citealt{nagel81_1},
\citealt{meszaros85}, 
\citealt{burnard91}, \citealt{becker05}).  However no 
self-consistent, general model applicable to X-ray sources has been 
established, due to the complexity of the physical processes that take place 
in the accretion column and in the magnetosphere. The radiation spectrum 
or ``standard'' X-ray continuum \citep{white83} is a power law in 
the $\sim(5-20)\,$keV energy range with an exponential cutoff at energies  
$\sim(20-30)\,$keV  \citep{coburn02}. Additionally, iron fluorescence 
Fe K$\alpha$ lines are produced in the circumstellar material, and 
cyclotron absorption lines also modify the continuum emission (see Sec.~\ref{sec:crsf}). 

\begin{figure}[h]
\resizebox{\hsize}{!}{\includegraphics[clip=true]{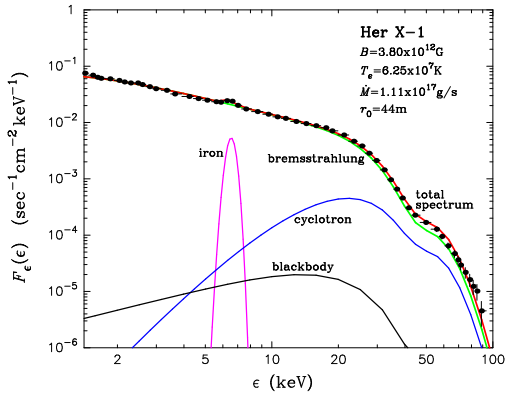}}
\caption{
\footnotesize
Theoretical spectrum of Her\,X-1. The total spectrum (red line) and 
individual components are shown, together with a \textsl{BeppoSAX} spectrum of the source 
(black circles). (Figure 6 from \cite{becker_wolff_07}, 
reproduced by permission of the AAS).}
\label{fig:bw_6}
\end{figure}

Recently, \cite{becker_wolff_07} have developed a 
model for the continuum formation to reproduce the spectra of bright accreting X-ray 
pulsars, for which a radiative shock located above the stellar surface dominates the 
spectral formation. The total observed spectrum is a result of the bulk and thermal Comptonization 
of bremstrahlung, black body and cyclotron seed photons. A schematic view of the model is 
shown in Fig.~\ref{fig:bw_1}. In Fig.~\ref{fig:bw_6} a comparison between 
the theoretical spectrum of Her\,X-1 and a \textsl{BeppoSAX} observation of the source is shown.  

\begin{figure}[]
\resizebox{\hsize}{!}{\includegraphics[clip=true]{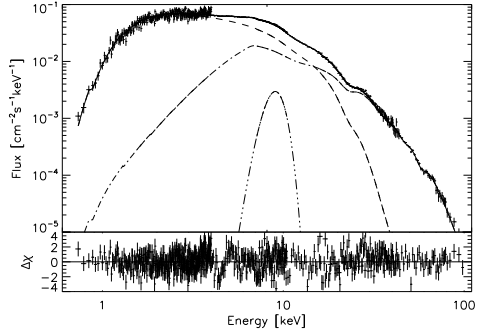}}
\caption{
\footnotesize
Unfolded spectrum of 4U\,0115+64 and best fit model. The solid line shows 
the best fit model, sum of the dashed line that represents the thermal Comptonization, the dot-dashed line 
that shows the column emission, and the dot-dot dashed line that shows a narrow Gaussian line. Lower 
panel: residuals from the best fit model. 
From \cite{ferrigno09}, reproduced with permission \textcopyright~ESO.}
\label{fig:ferrigno}
\end{figure}

The model is available for spectral fitting in \textsl{XSPEC} \citep{ferrigno09}. The authors have 
applied the model to \textsl{BeppoSAX} observations of the accreting Be/X-ray binary 4U\,0115+64 
during an outburst. From this study, Ferrigno et al. found that the emission above $\sim$7\,keV could 
be due to thermal and bulk Comptonization of the seed photons produced by cyclotron cooling of the 
accretion column, plus emission at lower energies due to thermal Comptonization of a blackbody 
component from a diffuse halo close to the neutron star surface. The data, best fit model, and residuals 
are shown in Fig.~\ref{fig:ferrigno}.    

\section{Cyclotron Resonance Scattering Features}\label{sec:crsf}

\begin{table*}
  \caption[List of cyclotron sources]
          {List of sources with cyclotron line(s) significantly detected 
            in their spectrum, updated from \cite{makishima99} and  
            \cite{heindl04}. The cyclotron line energy, the discovery 
            instrument and reference are given. For the last five sources 
            listed there is weak evidence for a cyclotron line but this has 
            not been firmly confirmed. }
   \vspace{-0.2cm}
  \label{tab:crsf} 
  \centering
  \begin{tabular}{|llll|}\hline
Source & $E_{\mathrm{n,cyc}}\,$(keV) & Discovery        & Reference           \\\hline
Swift J1626.6-5156 &10             &\textsl{RXTE}     & \cite{deCesar09}    \\
4U\,0115+64        &14,24,36,48,62 &\textsl{HEAO-1}   &\cite{wheaton79}     \\
4U\,1907+09        &18,38          &\textsl{Ginga}    &\cite{makishima92}   \\
4U\,1538-52        &20             &\textsl{Ginga}    &\cite{clark90}       \\
Vela\,X-1          &24,52          &\textsl{HEXE}     &\cite{kend92}        \\
V\,0332+53         &26,49,74       &\textsl{Ginga}    &\cite{makishima90}   \\
Cep\,X-4           &28             &\textsl{Ginga}    &\cite{mihara91}      \\
Cen\,X-3\          &28.5           &\textsl{RXTE}     &\cite{heindl99_1}    \\
                   &               &\textsl{BeppoSAX} &\cite{santangelo98}  \\
X Per\             &29             &\textsl{RXTE}     &\cite{coburn01}      \\
MXB\,0656-072      &36             &\textsl{RXTE}     &\cite{heindl03}      \\
XTE\,J1646+274     &36             &\textsl{RXTE}     &\cite{heindl01}      \\
4U\,1626-67        &37             &\textsl{RXTE}     &\cite{heindl99_1}    \\
                   &               &\textsl{BeppoSAX} &\cite{orlandini98}   \\
GX\,301-2          &37             &\textsl{Ginga}    &\cite{makishima92}   \\
Her\,X-1\          &41             & Balloon          &\cite{truemper77}    \\
A\,0535+26         &46,100         &\textsl{HEXE}     &\cite{kend92}        \\
1A\,1118-61        & 55            & \textsl{RXTE}    &\cite{doroshenko2010}\\
EXO\,2030+375      &11?            &\textsl{RXTE}     &\cite{wilson08}      \\
                   &63?            &\textsl{INTEGRAL}              &\cite{klochkov08_2}  \\
GS\,1843+00        &20?            &\textsl{Ginga}    &\cite{mihara95}      \\
OAO\,1657-415      &36?            &\textsl{BeppoSAX} &\cite{orlandini99}   \\
GRO\,J1008-57      &88?            & \textsl{CGRO}    & \cite{shrader99}    \\
LMC\,X-4\          &100?           &\textsl{BeppoSAX} &\cite{labarbera01}   \\\hline
  \end{tabular}
 \vspace{-0.3cm}
\end{table*}

The energy of the electrons in the highly magnetized plasma is quantized into 
Landau levels. Due to the resonant scattering of photons off electrons, 
absorption-like features can be formed in the X-ray spectrum 
of X-ray pulsars (cyclotron resonance scattering features or simply cyclotron 
lines). The energy spacing between the Landau levels is given by

\begin{equation}
E_{\mathrm{cyc}}= \hbar\frac{eB}{m_{\mathrm{e}}c}=11.6\,\mathrm{keV}\cdot B_{12}
\end{equation}
where $B_{12}$ is the magnetic field in units of $10^{12}$ G. Therefore, 
the measurement of a cyclotron line in the spectrum of a highly magnetized 
accreting neutron star provides a direct measurement of the neutron star's 
magnetic field. Due to the strong gravitational field around 
the neutron star, the cyclotron line energy is gravitationally red-shifted, 
and the observed line energies have to be corrected as 
\begin{equation}
E_{n}=nE_{\mathrm{cyc}}=(1+z)E_{n,\mathrm{obs}}
\end{equation}
where the gravitational redshift is $z\sim1.25\dots1.4$ for typical neutron star parameters. 

Since the discovery of a cyclotron line in the X-ray spectrum of Her X-1 in 
1976 \citep{truemper78}, cyclotron lines have been observed in 16 
accreting pulsars (see e.g. \citealt{makishima99}, \citealt{heindl04}, 
for reviews, and Table~\ref{tab:crsf} for a list of the currently known sources 
that exhibit cyclotron lines). Several sources exhibit a fundamental 
line plus harmonics, like V\,0332+53 \citep{kreykenbohm05}. The record holder is 
4U\,0115+64, showing up to five cyclotron lines in its spectrum (\citealt{santangelo99_2}, 
\citealt{heindl99_2}). 

Cyclotron lines are extremely powerful tools 
since they provide the \emph{only} direct way to determine the magnetic 
field of a neutron star, and they can also probe the change of the accretion 
structure and plasma properties with luminosity. 

Matter is channeled onto the magnetic poles by the magnetic field lines
creating accretion funnels, where the X-ray emission is originated. As 
proposed in the pioneering work of \citet{basko_sunyaev_76}, there are two 
different accretion regimes according to the mass accretion rate (and hence 
the luminosity). For luminosities higher than a critical luminosity, which is 
generally comparable to the Eddington luminosity $L_{\mathrm{E}}$ \citep{nelson93}, 
a radiation dominated shock is formed in the accretion column. In this case 
a ``fan beam'' emission pattern is predicted,  in which photons escape 
perpendicular to the accretion column. For luminosities lower than the critical 
luminosity, matter can fall almost freely until the neutron star surface, causing 
the braking of the plasma by a hydrodynamical shock close to the neutron star surface. 
In this case, X-rays are able to escape vertically along the accretion column, 
giving rise to a ``pencil beam'' pattern.
\begin{figure}[]
\resizebox{\hsize}{!}{\includegraphics[clip=true]{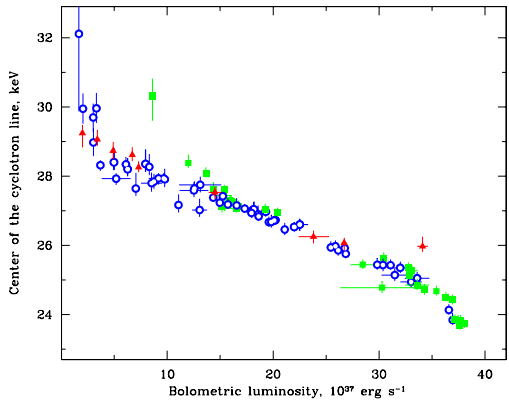}}
\caption{
\footnotesize Evolution of the cyclotron line energy of V\,0332+53 with the X-ray luminosity. From 
\cite{tsygankov2010}, reproduced with the permission from MNRAS. }
\label{fig:0332}
\end{figure}
For high luminosity sources it is generally believed that an increase in the 
mass accretion rate (or increase in luminosity) causes an increase in the 
height of the accretion column (\citealt{burnard91}, based on 
\citealt{basko_sunyaev_76}). Assuming a dipole magnetic field, the height of 
the accretion column can be estimated in terms of the cyclotron energy. 
Therefore, cyclotron line energy measurements allow us to test this theory. 
This relation was experimentally confirmed with \textsl{Ginga} observations 
of 4U\,0115+64  (\citealt{mihara95}, \citealt{mihara04}, \citealt{nakajima06})
 and \textsl{RXTE} observations of 4U\,0115+64 (\citealt{nakajima06}, \citealt{tsygankov2010}) 
and V\,0332+53 \citep{tsygankov07}. For these sources, a negative correlation 
between the X-ray luminosity and the cyclotron line energy has been observed 
(see Fig.~\ref{fig:0332}).

\begin{figure}[]
\resizebox{\hsize}{!}{\includegraphics[clip=true]{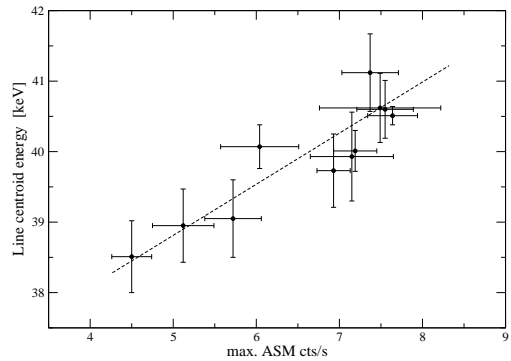}}
\caption{
\footnotesize 
Dependence of cyclotron line energy with the X-ray luminosity for Her\,X-1. 
From \cite{staubert07}, reproduced with permission \textcopyright~ESO.}
\label{fig:herx1_staubert07}
\end{figure}

For low luminosity sources in the sub-Eddington regime, a different
behavior is expected. \cite{staubert07} have discovered for 
Her\,X-1 a positive correlation between the cyclotron line energy and the 
luminosity (see Fig.~\ref{fig:herx1_staubert07}). They find a change in the 
cyclotron line of 5\% for a luminosity variation of $\Delta L/L\sim 1$.
In this case, no radiation dominated shock is expected to 
form in the accretion column. Based on \cite{basko_sunyaev_76} and 
\cite{nelson93}, where the physics of accretion in low-luminosity sources 
is studied, \cite{staubert07} find the following relation between the 
cyclotron line energy and the luminosity:
$\Delta E_{\mathrm{cyc}}/E_{\mathrm{cyc}}=3\frac{l_{\ast}}{R}\frac{\Delta L}{L}$
where $l_{\ast}$ is the height of the scattering region above the neutron star
surface and $R$ is the neutron star radius. For $\Delta L/L\sim 1$, 
this implies a change of $\sim$3\%, very close to the 5\% obtained from the 
Her\,X-1 observations.

Contrary to the above sources, measurements of the 
cyclotron line energy of A\,0535+26 during a normal outburst in 2005 show 
no significant correlation between the cyclotron line energy and the luminosity. This 
suggests that the line forming region does not change with the mass accretion rate of the system 
\citep{caballero07}. A question that is still under debate is whether the transition between the 
two regimes has been observed.

\begin{figure}[]
\resizebox{\hsize}{!}{\includegraphics[clip=true]{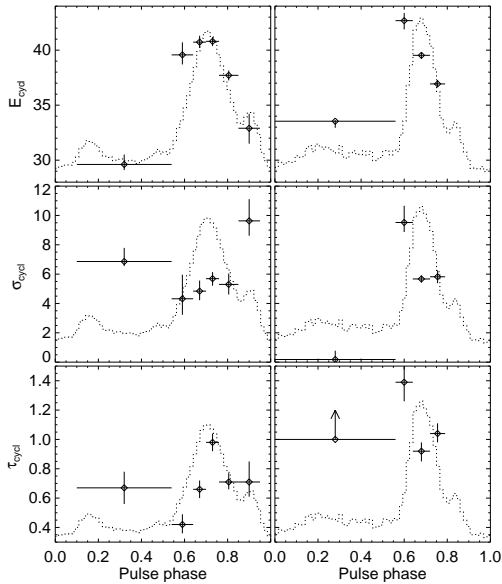}}
\caption{
\footnotesize Variation of the cyclotron line parameters with the pulse phase for Her X-1, during the start
of the main-on state (\textsl{left}) and at the end of the main-on state (\textsl{right}). From 
\cite{klochkov08_1}, reproduced with permission \textcopyright~ESO.} 
\label{fig:ecyc_herx1}
\vspace{-16.pt}
\end{figure}

Changes in the cyclotron line energy have been reported not only with the luminosity, but also with the 
pulse phase. The cyclotron line energy values discussed above correspond to phase averaged spectra. However, 
due to the rotation of the neutron star, during a rotational phase the observer is looking into different 
regions of the neutron star surface and/or accretion column, and therefore different 
magnetic fields and continuum parameters can be expected. For GX\,301-2, 
\cite{kreykenbohm04_2} have shown that 
the cyclotron line energy varies by 25\%, implying a variation of the magnetic field between 
$3.4\times10^{12}$ and $4.2\times10^{12}$\,G. For Her X-1, \cite{klochkov08_1} have also found a variation 
of the cyclotron line and continuum parameters with the pulse phase (see Fig.~\ref{fig:ecyc_herx1}), 
that they interpret with different viewing angles of the neutron star along the rotational phase. 

\begin{figure}
\resizebox{\hsize}{!}{\includegraphics[clip=true]{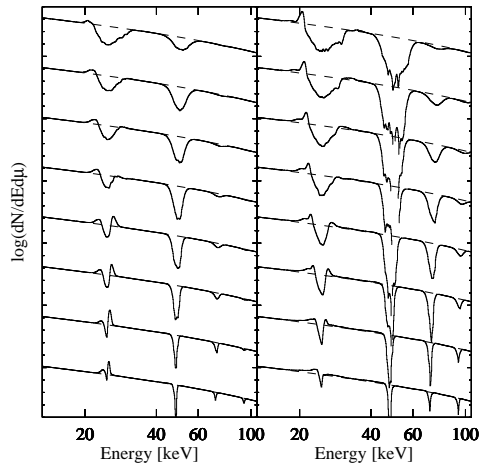}}
\caption{
\footnotesize
Simulated cyclotron line shapes for $B/B_{\mathrm{crit}}=0.04$ as a function
of angle (decreasing upwards) and optical depth. A cylindrical geometry, a plasma temperature of 
$kT=3$\,keV,
Thomson optical depths of $3\times10^{−4}$ (left) and $3\times10^{−3}$ (right) are assumed. Line shapes are rather complex
and strongly depend on the physical and geometrical parameters. From \cite{schoenherr07}, reproduced with permission \textcopyright~ESO.}
\label{fig:cyclo}
\end{figure}

Until recently, only phenomenological models (Gaussian or Lorentzian 
curves) were used to describe the cyclotron lines observed in the spectra 
of accreting X-ray pulsars. However, Monte Carlo simulations of the 
propagation of photons through a low density plasma assumed to be threaded 
by an uniform magnetic field have revealed that the cyclotron lines are 
expected to vary in shape, depth and width over the pulse  
(\citealt{araya99}, \citealt{araya00}, \citealt{schoenherr07}, see  
Fig.~\ref{fig:cyclo}). A complex line profile has also emerged from 
observations (e.g., in V\,0332+53, \citealt{pottschmidt05}). A Monte Carlo code 
based on the code of \cite{araya99}, \cite{araya00} has been implemented in 
\textsl{XSPEC} by \citep{schoenherr07}. First fits 
have been performed to Cen X-3 data \citep{suchy08}. The new model allows  
to narrowly constrain the main physical parameters of the system, as the 
plasma geometry (slab versus cylinder), the electron parallel temperature, 
and density. The latest version of the model 
allows for complex geometries, magnetic field gradients, and velocity gradients 
\citep{schwarm2010}.

\section{Conclusions}

We have reviewed the general properties of accreting X-ray pulsars, in particular the spectral 
formation and cyclotron lines. Many aspects of accreting X-ray pulsars have not been 
discussed in this review, like pulse profile formation. Modeling of pulse profiles has been 
performed for instance by \citet{wang81}, \citet{meszaros85}, and \cite{leahy91}.
An alternative pulse profile decomposition method has been proposed by \cite{kraus95}, that
allows to infer geometrical parameters of the neutron star, and to reconstruct the emission beam pattern 
that can be compared to model predictions (see \cite{kraus03} and e.g. \citealt{sasaki10}). Also not discussed 
in this review are the relativistically broadened lines that have been observed in several neutron star X-ray binaries.  
These can be used to constrain neutron star parameters and the equation of state, see, e.g.,  \cite{cacket08} but also \cite{done2010}. 

Forty years after the discovery of the first accreting neutron star, the continuum formation is now better understood, 
but still many questions are open. There are now 16 neutron stars with direct magnetic field determinations -- now enough 
data available for both, individual studies, and study as a class. The cyclotron line behavior with the X-ray luminosity allows 
deeper probing of accretion column theory. Good numerical models for cyclotron line formation are available, and the behavior 
of the lines is roughly in agreement with predictions of Monte Carlo computations. 

In the future, \textsl{MAXI} \citep{matsuoka09} and Swift/BAT \citep{barthelemy05} 
will continue to monitor the X-ray sky and discover neutron star outbursts. 
\textsl{ASTROSAT} \citep{OBrien2011} will allow routine measurements of broadband spectral shape and 
magnetic fields, \textsl{NuStar} \citep{harrison05} will start to resolve the shape of cyclotron lines, while 
\textsl{LOFT} \citep{feroci10} and \textsl{Athena} will allow studies at characteristic variability timescale. 
With the current and future missions, we have reached a very exciting era in which high quality data are available 
together with physical models that will allow a deeper understanding of the extreme physics involved in the accretion of 
matter onto a neutron star. 
\begin{acknowledgements} 

\footnotesize{IC acknowledges financial support from the French Space Agency CNES through CNRS, and Andrea  
Santangelo for the help in the organization and the strong support. 
The authors acknowledge Franco Giovannelli and organizers for the invitation to the conference, 
and thank the MAGNET collaboration and Katja Pottschmidt for her useful comments on the manuscript.
JW acknowledges partial support by DLR grant 50 OR 1007.}  

\end{acknowledgements}
\vspace{-0.6cm}
{\small \bibliographystyle{aa}

}
\end{document}